\documentclass[twocolumn,showpacs,amsmath,amssymb]{revtex4}
\usepackage{amsmath}
\usepackage{amssymb}
\usepackage{graphicx}
\usepackage{dcolumn}
\usepackage{bm}

\begin{document}
\title{Active Motion Assisted by  Correlated Stochastic  Torques}

\author{Christian Weber$^1$ $^2$, Paul K. Radtke$^1$, Lutz
  Schimansky-Geier$^1$\email{{alsg@physik.hu-berlin.de}}, and
  Peter H\"anggi$^3$ }
\affiliation{$^1$Institute of Physics, Humboldt University at Berlin,
  Newtonstr. 15, D-12489 Berlin, Germany\\$^2$INRIA Nancy-Grand Est, Team CORTEX, 615 Rue du Jardin Botanique,
54600 Villers-Les-Nancy, France\\$^3$Institute of Physics,
  University of Augsburg, Universit\"atsstra{\ss}e 1, D-86135
  Augsburg, Germany}

\begin{abstract}
  The stochastic dynamics of an active particle undergoing a constant speed
  and additionally driven by an overall fluctuating torque is
  investigated. The random torque forces are expressed by a stochastic
  differential equation for the angular dynamics of the particle
  determining the orientation of motion. In addition to a constant
  torque, the particle is supplemented by random torques which are
  modeled as an Ornstein-Uhlenbeck process with given
  correlation time $\tau_c$.  These nonvanishing correlations cause a
  persistence of the particles' trajectories and a change of the
  effective spatial diffusion coefficient. We discuss the mean square
  displacement as a function of the correlation time and the noise
  intensity and detect a nonmonotonic dependence of the effective
  diffusion coefficient with respect to both correlation time and
  noise strength. A maximal diffusion behavior is obtained if the
  correlated angular noise straightens the curved trajectories,
  interrupted by small pirouettes, whereby the correlated noise
  amplifies a straightening of the curved trajectories caused by the
  constant torque.
\end{abstract}
\maketitle

\section{Introduction}
Recent experimental studies evidenced that living beings, such as
bacteria \cite{beta}, different water fleas \cite{daphniapattern},
fish \cite{fish}, birds \cite{edwards} and insects
\cite{bazazi,BazRom}, are able to sustain a constant mean speed over
large time scales. Thus, these animals do not only perform passive Brownian
motion, but instead exhibit a so-called active stochastic motion
\cite{SchEb98}, occurring far from equilibrium.

Such systems form a large field of recent theoretical interest
\cite{MiCa01,schweitzer}, including the influence of fluctuations. One
might relate the sources of fluctuations on animal motion either in the
food supply \cite{StEb09} or in the propulsive engine \cite{RoLsg11}
that drives the unit or by assuming a molecular agitation imposed externally
outside \cite{beta}. In this approach one widely uses similarities of
the animal motion and Brownian motion which Einstein and Smoluchowski
independently described in a probabilistic theory
\cite{einstein,smoluchowski}.  Later on, Langevin encoded this theory
of stochastic systems into the Newtonian notation of equations of
motion \cite{langevin}, which present a first formulation of a
stochastic differential equation.  This concept, completed by an active
element describing the self-propulsion, is nowadays widely used for the
description of fluctuating animal motion \cite{MiCa01,schweitzer,SchEb98,StEb09,RoLsg11}. Of foremost
interest are the specific effects of the random impacts on the motion.

Our focus here is on agents subjected to a torque, resulting in circular
motions. This feature can be caused, for example, by an external
magnetic field with profound physiological implications
\cite{kanokov}, or due to boundary conditions \cite{fish}. The agents
themselves may exhibit a preferred turning direction, either caused by
asymmetries in the propulsion as occurring in the chemotaxis of sperm
cells \cite{friedrich2,sperm2}, as a search strategy
\cite{WeSr81,MuWe94}, or also by interaction with other agents, which
leads to the emergence of swarming characteristics
\cite{daphniapattern,pavelswarm}.

We start out by studying the planar motion of active particles
with a constant velocity $v_0$ subjected to a constant torque
$\Omega$. To take into account thermal or, more generally, the
stochastic influence of the surrounding, we first consider an
additional white Gaussian noise $\xi(t)$ driving the angle dynamics
and discuss the resulting effective diffusion coefficient
$D_{\text{eff}}$ \cite{lsgzigzag,TefLow08,HaLsg08,TefLo09}. Next, we
propose a more realistic model considering some persistence in the
motion. Therefore, we replace the white Gaussian noise by a
time-correlated Gaussian noise (colored noise), namely an
Ornstein-Uhlenbeck process (OUP) $\theta(t)$
\cite{uhlenbeck,Horsthemke_Lefever,PH78,HMG84,Hwa89,JunHan_color},
which implies a memory for the curvature feature.  This generalization
then allows applications reaching from stochastic polymer dynamics
\cite{doi} over spiral wave motion \cite{SenAl00} up to animal
trajectories \cite{fish}. This present study using a constant torque
$\Omega$ can principally also be applied to a different problem,
namely the control of diffusion of electron beams in drift chambers
in presence of additional magnetic fields \cite{blum}. We want to
mention that other noise statistics, especially white Poissonian noise
\cite{hanggi,JunHan_color,RoLsg11}, are conceivable. Their study would
form an interesting future extension of the present work.
 
In a driven motion with a constant speed $v_0$ the correlations with
correlation time $\tau_c$ generate a persistence length $l_c$ of the
trajectory: $l_c=v_0 \tau_c$. As will be shown below, the inclusion of
correlations will hence characteristically modify the dynamics. The
resulting dynamics will be quantified by the spatial diffusion
coefficient $D_{\rm eff}$.

\section{Agents driven by white Gaussian noise}
\label{white} Our starting point is the equation of motion in
two dimensions with a constant velocity $v_0$ and a random torque with the mean value
$\Omega$. The dynamics for the position vector ${\vec
{r}}(t)=\{x(t),y(t)\}$ derives from its   velocity vector, reading
\begin{align}
{\rm {d \over dt}}\vec{r}(t)\,= \, v_0 \vec{e}_{v(t)}= \, v_0
\,\Big(\cos{\phi(t)},\, \sin{\phi(t)} \Big) \,, \label{whitesysr}
\end{align}
with  $\phi(t)$ denoting the orientation of the velocity
vector. This orientation $\phi(t)$, if governed  by
a constant torque and supplemented by random fluctuations,
yields a stochastic dynamics for $\phi(t)$
\begin{align}
  {\rm {d \over dt}} {\phi}(t)\,=\, \Omega + \frac{\sqrt{2D_{\xi}}}{v_0}\xi(t)\,.
\label{whitesys}
\end{align}
We consider the noise $\xi(t)$ to be a white Gaussian noise with
the corresponding noise intensity denoted by $D_{\xi}$ and with vanishing mean. The
$1/v_0$ in front of the random force is caused by the fact that the
tangential acceleration scales as $v_0 \dot{\phi}$. It expresses the
circumstance that fast agents cannot change their direction as quickly
as slower ones.

For a vanishing noise intensity the constant torque leads to circular
shaped motion, whose size is dictated by their cyclotron radius $R =
v_{0} / \Omega$. A non-vanishing noise intensity on the other hand
induces a more erratic behavior. The mean square displacement (MSD)
corresponding to this dynamics can be directly calculated by using
the shifted Gaussian transition probability distribution $P(\Delta
\phi,\tau)$. Then  the general expression for the MSD of a particle with a constant velocity $v_0$, i.e.
\begin{equation}
 \left< (\vec{r}(t)-\vec{r}_0)^2\right> =2v_0^2\int_0^t (t-\tau) \left<\mathrm{cos}\left(\Delta \phi(\tau)\right)\right>\mathrm d
\tau ,
\label{generalmsd}
\end{equation}
yields in the case of white Gaussian noise
\begin{equation}
  \left< (\vec{r}(t)-\vec{r}_0)^2\right>_\text{white}
= 2v_0^2\int_0^t (t-\tau) e^{-\frac{D_{\xi}\tau}{v_0^2}} \cos{(\Omega \tau)}\mathrm d\tau .
\label{whitemsd}
\end{equation}
For   small times $t$ a ballistic behavior $\propto t^2$ results while for
large times a crossover to diffusive motion $\propto t$ takes place. The
effective diffusion coefficient $D_{\text{eff}}$ in two dimensions is
related to the MSD via the well-known relation:
\begin{equation}
 D_{\text{eff}}=\lim_{t\rightarrow\infty}\frac{\langle (\vec{r}(t)-\vec{r}_0)^2\rangle}{4t}\,.
\label{deff}
\end{equation}
As a consequence we integrate Eq. (\ref{whitemsd}) in the limit of
large times. Furthermore, we substitute a dimensionless integration variable
$x= D_{\xi}v_{0}^{-2} \tau$  and obtain together with the new parameters
$D_{0}=v_{0}^{4}/(2D_{\xi})$ and $\omega_0=\Omega v_{0}^{2}/D_\xi$ an expression
for the effective diffusion coefficient, reading
\begin{align}
D_{\text{eff}}^{\text{white}} &= D_{0} \int_{0}^{\infty}e^{-x}\cos\left( \omega_0 x \right) dx= \frac{D_{0}}{1+\omega_0^{2}} \\&= \frac{D_{\xi}}{2\left[\left(\frac{D_{\xi}}{v_{0}^{2}}\right)^{2}+\Omega^{2}\right]}.
\label{deff_white}
\end{align}
Thus, we recover the result for a white Gaussian angle drive with
constant torque. This finding was  obtained before in Ref.
\cite{lsgzigzag} and has been re-addressed with Ref. \cite{TefLow08}, see
for more detailed discussions \cite{HaLsg08}. Obviously, the
diffusion as a function of the angular noise exhibits a maximum, with
$D_{\text{eff}}=v_0^2/(4\Omega)$ at $D_{\xi} = v_{0}^2 \Omega$, and
vanishes for both $D_{\xi}=0$ and $D_{\xi} \rightarrow \infty$. This
is contrasted to the behavior of the diffusion without a torque,
$D_{\text{eff}}^{\Omega=0}=v_{0}^4/(2D_{\xi})$
\cite{othmer,mikhailov}, which diverges for $D_{\xi}\rightarrow 0$.

The explanation is as follows: for small noise intensities, agents
subjected to a constant torque move in circles around a fairly
quasi-stationary center, whereas in absence of a torque the propagation proceeds along
almost straight lines. With noise included, both motion types
perform diffusion. This leads to a reduction of the mean square
displacement in case of zero torque. In contrast, agents subjected
 to a constant torque start to spread over the space which is expressed
by the numerator in Eq. (\ref{deff_white}), which grows linearly in
$D_\xi$. For large noise intensities $D_{\xi}$, this  leads to a suppression of
diffusion as accounted for in \eqref{deff_white}.
\begin{figure}[h]
\begin{center}
\includegraphics[width=0.95\linewidth]{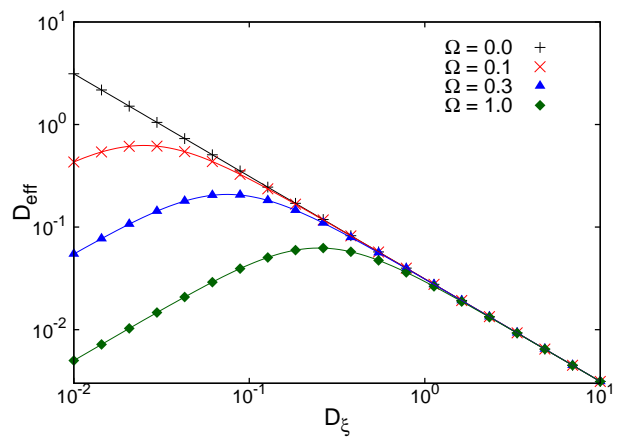}
\end{center}
\vspace{-10pt} \caption{(Color online) Effective diffusion coefficient $D_{\text{eff}}^{\text{white}}$ versus
  intensity $D_{\xi}$ of the Gaussian white noise for different torque strengths $\Omega$, with the constant velocity $v_0=0.5$ and $D_{\xi}=0.01$ within theory, see in  eq.
  (\ref{deff_white}).}
\label{phiwhitefordeff}
\end{figure}

We note that a finite torque always decreases
$D_{\text{eff}}^{\text{white}}$, because the cosine-contribution in
Eq. (\ref{whitemsd}) suppresses the value of the integral. This
difference becomes most striking for small noise intensity $D_{\xi}$.
In Fig. \ref{phiwhitefordeff} the behavior of eq. (\ref{deff_white})
is depicted as a function of $D_{\xi}$ for several $\Omega$-values.


\section{Agents Driven by an Ornstein-Uhlenbeck-Process}
\subsection{General outlines}
As a physically more realistic extension of our model, we next study time-correlated noise (i.e. {\it colored noise})
for the  angular drive instead of delta-correlated noise around its
mean  $\Omega$. Accordingly, we use an OUP as the most natural
continuous-valued colored noise
\cite{Horsthemke_Lefever,PH78,HMG84,Hwa89,JunHan_color}.  Hence, our new
system with the constant velocity $v_0$ and the torque $\Omega$ is
described by the modified angle dynamics, reading
\begin{align}
  \dot{\phi}(t)\,=\;& \Omega + \frac{1}{v_0} \theta(t) ,
  \label{gyrosysang}\\
  \dot{\theta}(t)\,=\;&-\frac{1}{\tau_c} \theta(t) +
  \sqrt{\frac{2D_{\xi}}{\tau_c^n}}\xi(t).
\label{gyrosys}
\end{align}
This colored noise dynamics assumes one additional auxiliary
process, namely the OUP $\theta(t)$ with the correlation time
$\tau_c$, the Gaussian white noise $\xi(t)$ and the corresponding
noise intensity $D_{\xi}/\tau_c^n$.

Note that we  included here a dependence of the correlation time to
the power $n$ in the noise intensity implying  a corresponding
change in dimension for  $D_\xi$ as $n$ changes. In doing so,
several different physical features can be addressed with no need to
perform additional calculations. For example, for $n=0$ the random
variable $\theta(t)$ merges with the Wiener process $W(t)$ in the
limit of infinite correlation times $\tau_c \to \infty$. In specific
detail: $\dot \theta =\sqrt{2D_{\xi}}\xi(t) \Rightarrow
\theta(t)=\sqrt{2D_{\xi}}\int^{t} \xi(t^\prime)
dt^\prime=\sqrt{2D_{\xi}} W(t)$. On the other hand,
the case with $n=1$ delivers a variance of $\theta$,
which is independent of the correlation time \cite{JunHan_color}.

>From a theoretical point of view, the case with $n=2$ is of utmost
interest. The limit $\tau_{c}\rightarrow 0$ yields Gaussian white
noise, i.e. $\theta(t) \rightarrow \sqrt{2D_{\xi}}  \xi(t)$, and thus allows a comparison with  the findings taken from
Sec. \ref{white}.

The corresponding $\theta$-correlation function reads in the
stationary limit
\begin{equation}
  \langle \theta(t)\theta(0)\rangle_s\,=\;\tau_c^{1-n}D_{\xi} e^{-\frac{|t|}{\tau_c}} ,
  \label{gyrotcorr}
\end{equation}
depicting a strong $n$-dependence of the variance, i.e.  the factor in front of the exponential.

To study the angular dynamics $\phi(t)$, we use the results of
Ornstein and Uhlenbeck \cite{uhlenbeck}, which were later generalized
by Chandrasekhar to motion in a higher dimensional space
\cite{chandrasekhar}. Therewith we can write the probability
distribution in the stationary limit for a transition during $\tau$
from $\phi_1$ to $\phi_2$ and conditioned by $\theta_1$ as
\begin{multline}
 P_{\mathrm{OUP}}(\phi_2,\tau|\phi_1,\theta_1)\;=\sqrt{\frac{v_0^2}{2\pi \tau_c^{3-n} D(\tau)}}\\
\times \exp \left[-\frac{v_0^2\left(\phi_2-\phi_1-\Omega \tau -\theta_1\tau_c\left(1 - e^{-
\frac{\tau}{\tau_c}}\right)\right)^2}{2 \tau_c^{3-n} D(\tau)}\right].
\label{cpdchandra}
\end{multline}
Here, the mean square increment of the angle holds
\begin{equation}
 D(\tau)=D_{\xi} \left(\frac{2\tau}{\tau_c} -3 + 4 e^{-\frac{\tau}{\tau_c}}
- e^{-\frac{2 \tau}{\tau_c}}\right) .
\label{cpdchandraD}
\end{equation}
This resulting Gaussian expression can be directly used to calculate
the mean square displacement of our angle dynamics from Eq.
(\ref{generalmsd}), yielding with statioary Gaussian distributed
initial values $\phi_1$ and $\theta_1$
\begin{equation}
 \langle (\vec{r}(t)-\vec{r}_0)^2\rangle_{\text{OUP}}=\;2v_0^2\int_0^t(t-\tau)
 e^{-\Psi(\tau)} \cos{(\Omega \tau)}\mathrm d\tau \,.
\label{msd}
\end{equation}
The function in the exponent reads
\begin{equation}
 \Psi(\tau) =\;\Psi_0\left(\frac{\tau}{\tau_c} - 1 + e^{-\frac{\tau}{\tau_c}}\right) .
\end{equation}
wherein we define
\begin{equation}
  \label{eq:psi0}
 \Psi_0\,=\, \frac{\tau_c^{3-n} D_{\xi}}{ v_0^2 }\,,
\end{equation}
as a parameter which scales monotonously with $\tau_c$ and $D_\xi$ and
can therefore be used to discuss the asymptotic behavior, later on.

Without constant torque ($\Omega =0$) and with $n=0$ this reproduces
the result of \cite{fish2}, where the motion of fish was analyzed
within a similar model.  Furthermore, Eq. (\ref{msd}) exhibits a
crossover from a ballistic behavior $\propto t^2$ to diffusive motion
$\propto t$. This can be seen by studying the following two temporal
limits
\begin{align}
\lim_{t \to 0} {\rm {d \over dt}}
\langle (\vec{r}(t)-\vec{r}_0)^2\rangle_{\text{OUP}} &= 2\,v_0^2\,t\, ,\\
\lim_{t \to \infty} {{\rm d} \over { \rm d}t}
 \langle(\vec{r}(t)-\vec{r}_0)^2\rangle_{\text{OUP}}&= 2 v_0^2 \int_0^\infty e^{-\Psi(\tau)}\cos{(\Omega\tau)}{\rm d}\tau .
\label{tinfder}
\end{align}
The right-hand side of Eq. (\ref{tinfder}) is obviously a non-vanishing
constant which can be identified with the effective diffusion
coefficient via Eq. (\ref{deff}), yielding our central finding, namely
\begin{equation}
  D_{\text{eff}}^{\text{OUP}}= \frac{v_0^2}{2} \int_0^\infty e^{-\Psi(\tau)}\cos{(\Omega\tau)}{\rm d}\tau.
\label{deffoup_dim}
\end{equation}
For further analytical discussions it will be helpful to use a dimensionless representation of Eq. (\ref{deffoup_dim}). To this end, we substitute $x=\frac{\tau}{\tau_c}\Psi_{0}$ and introduce the rescaled variables
\begin{equation}
D_c=\frac{v_0^4}{2\tau_{c}^{2-n}D_{\xi}}	\qquad\text{and}\qquad	\omega_c=\frac{v_{0}^{4}\Omega}{\tau_{c}^{2-n}D_{\xi}},
\label{dcwc}
\end{equation}
yielding
\begin{equation}
   D_{\text{eff}}^{\text{OUP}}= D_c \int_0^\infty e^{-x}e^{\Psi_0\left(1-e^{-x/\Psi_{0}}\right)}\cos{(\omega_c x )}{\rm d}x.
\label{deffoup}
\end{equation}
In comparison with the results for a white Gaussian angle drive [c.f.  Eq.
(\ref{deff_white})] we notice different definitions of the constants
$D_c$ and $\omega_c$ and an additional exponential in the integrand,
namely $e^{\Psi_0\left(1-e^{-x/\Psi_{0}}\right)}$. The latter
converges to unity for $\Psi_{0}\rightarrow 0$ and increases
monotonously with $\Psi_{0}$.
For $n=2$, the parameters
$\omega_c$ and $D_c$ coincide with those defined in the white
noise case, respectively, with $D_0$ and $\omega_0$. In this case $\Psi_{0}$
describes the deviation from the results under white noise.

Also note that the integral in Eq. \eqref{deffoup} can be expressed in a serial
expansion. Substituting first $z=\Psi_{0}\exp(-x/\Psi_{0})$ as a new
variable and expanding afterwards the remaining exponential under the
integral in a Taylor series, one can calculate the integral in each
summand, yielding
\begin{equation}
  \label{seriesdimless}
  D_{\text{eff}}^{\text{OUP}}= D_c e^{\Psi_0} \sum_{k=0}^\infty\frac{(-1)^k}{k!}\Psi_0^{k+1}\frac{\Psi_0+k}{(\Psi_0+k)^2+(\Psi_{0}\omega_c)^2} .
\end{equation}

The asymptotic behavior of the diffusion coefficient can be readily derived: For vanishingly small $\Psi_0$,
the first item
in the sum with $k=0$ dominates. Hence we obtain
\begin{equation}
  \label{eq:deff_as}
  \lim_{\Psi_0 \to 0} D_{\text{eff}}^{\text{OUP}}\,=\,   \lim_{\Psi_0 \to 0}  \frac{D_{c}}{1+\omega_c^2} ,
\end{equation}
which in particular recovers the white noise result of eq. \eqref{deff_white} for the case $n=2$.

The opposite asymptotics of large $\Psi_0$ can be inspected by looking
at the original integral definition in Eq. \eqref{deffoup}, (see also
\cite{fish2}) for the case without an applied torque and $n=0$. In this limit the
major contributions to the integral stem from values at the lower
boundary $x \sim 0$. Upon expanding the double exponent until the second
order finds
\begin{equation}
  \label{eq:deff_as1}
  \lim_{\Psi_0 \to \infty} D_{\text{eff}}^{\text{OUP}}\,\propto \lim_{\Psi_0 \to \infty} D_c \sqrt{\Psi_0}\exp \left(- \frac{\Psi_0 \omega_c^2}{2} \right),
\end{equation}
which tends to zero much faster compared to the case without torque with
$\omega_c=0$.

\subsection{Asymptotic behavior in the absence of  torque: $\Omega=0$}

\begin{figure}
\begin{center}
  \includegraphics[width=0.95\linewidth]{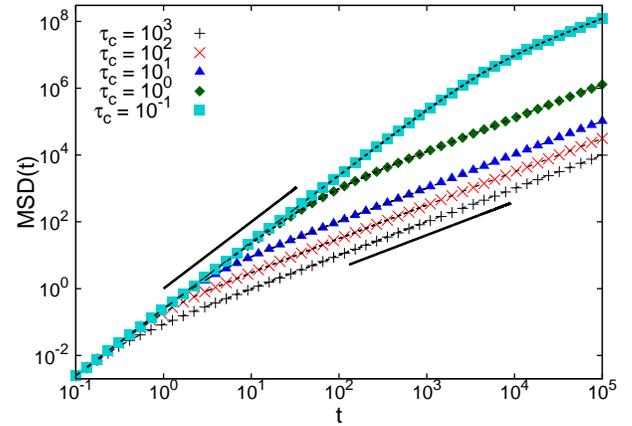}
\end{center}
\vspace{-10pt}
\caption{(Color online) Mean square displacement $\langle
  (\vec{r}(t)-\vec{r}_0)^2\rangle$ for the dynamics given with Eqs.
  (\ref{gyrosysang}) and (\ref{gyrosys}) for $\Omega=0$,
  $D_{\xi}=0.01$, $v_0=0.5$ and with $n=0$. Simulations (points) agree
  well with the theory (Eq. (\ref{msd}); dashed lines).
  The left bar is $\propto t^2$, the right bar is $\propto t$. For
  $\tau_c\gtrsim 3$, the ratio of subsequent crossover times behaves
  according to $t_\text{cross}(\tau_{c})/t_\text{cross} (10
  \tau_{c})=\sqrt{10}$, otherwise according to
  $t_\text{cross}(\tau_{c})/t_\text{cross} (10 \tau_{c})=10^2$.}
\label{oupvar}
\end{figure}

First we study  the crossover between the ballistic and the diffusive
behavior. Towards this aim we expand the two exponentials in Eq.
(\ref{msd}).  A case differentiation, due to the two corresponding
exponential time dependencies, yields two parameter dependent
regimes where the crossover between ballistic motion and diffusion is
realized
\begin{align}
   \mathrm{if}\quad \Psi_0\; < & \;1 \qquad \rightarrow \qquad
  t_{\mathrm{cross}} \approx \frac{v_0^2\tau_c^{n-2}}{D_{\xi}}\,,\\
   \mathrm{if}\quad \Psi_0\; > & \;1 \qquad \rightarrow \qquad
  t_{\mathrm{cross}} \approx \sqrt{\frac{v_0^2\tau_c^{n-1}}{D_{\xi}}}\,.
\label{cross2}
\end{align}
These crossovers are corroborated with our  simulation results and the
numerical evaluations of Eq. (\ref{msd}) in Fig. \ref{oupvar}. We see
that the ratio of crossover times where $\Psi_0>1$ holds (with the
assumed parameters this is true for $\tau_c\gtrsim3$), behaves
according to $t_\text{cross}(\tau_{c})/t_\text{cross}(10
\tau_{c})=\sqrt{10}$.  In case $\Psi_0<1$, the ratio of crossover
times increases severely to
$t_\text{cross}(\tau_{c})/t_\text{cross}(10 \tau_{c})=10^2$.

\begin{figure}
\begin{center}
\includegraphics[width=0.95\linewidth]{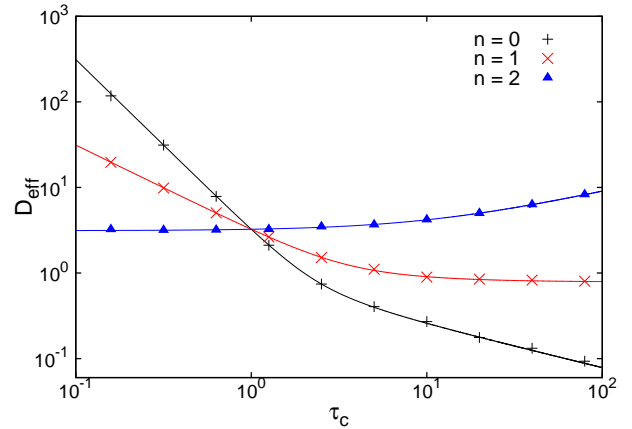}
\end{center}
\vspace{-10pt}
\caption{(Color online) The effective diffusion coefficients
  $D_\text{eff}$ are depicted as a function of the correlation time
  $\tau_c$ for several $n$ values within simulations (points) and
  theory (Eq. (\ref{msd}); lines). Other parameters as in Fig.
  \ref{oupvar}. While decreasing for $n=0$ and approaching a non-zero
  constant for $n=1$, $D_\text{eff}$ diverges $\propto \sqrt{\tau_c}$
  for $n=2$, see Eq. (\ref{deff_tc1}).  
  }
\label{tcorrdeff}
\end{figure}

\begin{figure}[b]
\begin{center}
\includegraphics[width=0.49\linewidth]{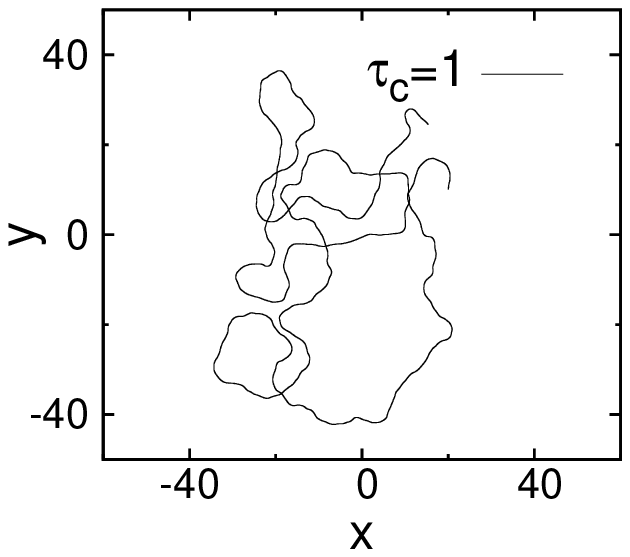}
\includegraphics[width=0.49\linewidth]{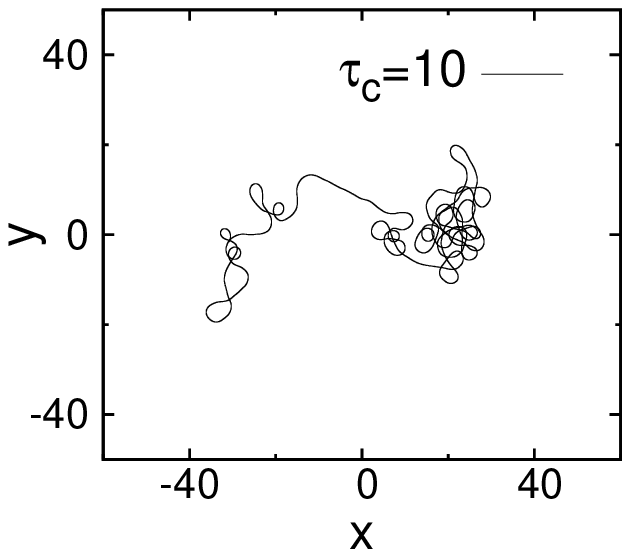}
\includegraphics[width=0.49\linewidth]{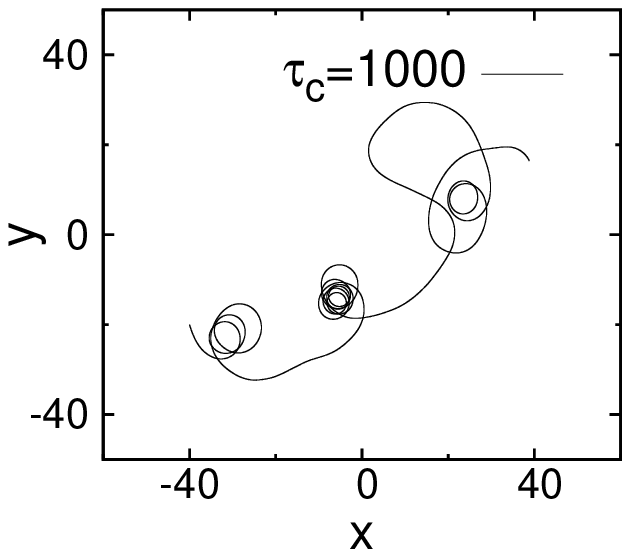}
\end{center}
\vspace{-10pt}
\caption{Spatial trajectory within an angular OUP dynamics with
  constant variance $n=1$ for different correlation times $\tau_c$
  while the parameters hold as in Fig. \ref{fortraj} but with
  $\Omega=0$, $n=1$ and the time length $t_l=1000$. The trajectories
  changes qualitatively from a random way being unoriented each moment
  to an agent performing a waltzing-like dance. The effective
  diffusion coefficient remains unchanged for the shown cases (cf.
  Fig. \ref{tcorrdeff}).}
\label{traj}
\end{figure}

The different crossover behaviors thus reflect the different scaling properties of
$D_\text{eff}$ for the different regimes of $\Psi_0$ in agreement with the result that
\begin{align}
  \mathrm{if}\quad \Psi_0\; < & \;1 \qquad \rightarrow \qquad
  D_{\text{eff}}\;\approx \frac{v_0^4\tau_c^{n-2}}{2 D_\xi},\label{deff_tc1}\\
\mathrm{if}\quad \Psi_0\; > & \;1 \qquad \rightarrow \qquad
D_{\text{eff}}\;\approx\; \frac{v_0^3}{2}\sqrt{\frac{\tau_c^{n-1}}{D_\xi}}\,.\label{deff_tc2}
\end{align}
\begin{figure}[t]
\begin{center}
\includegraphics[width=0.49\linewidth]{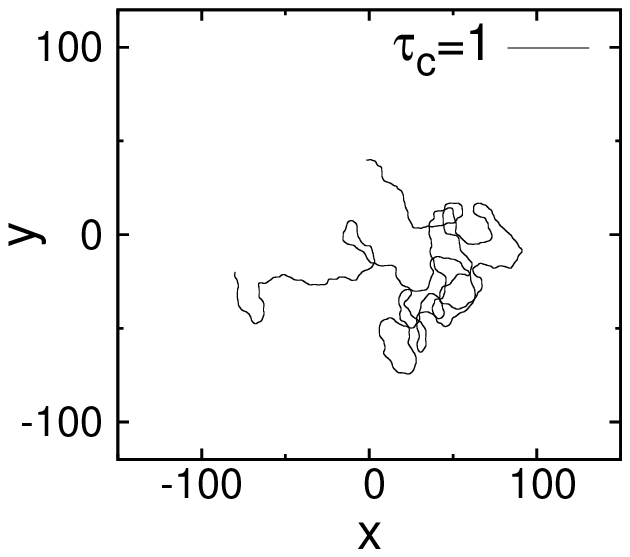}
\includegraphics[width=0.49\linewidth]{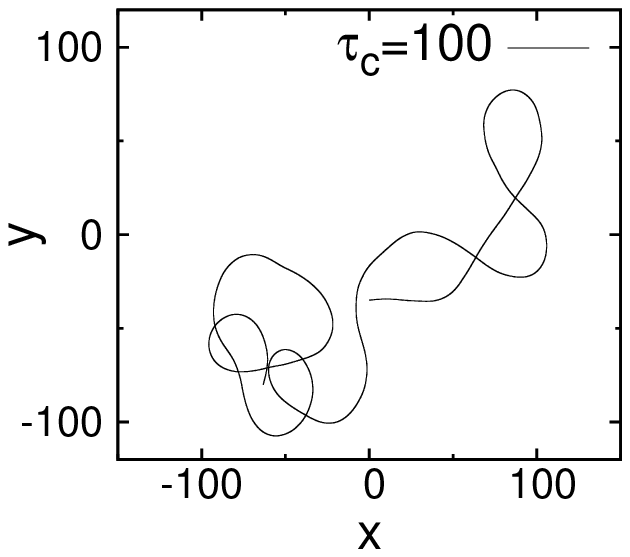}
\end{center}
\vspace{-10pt}
\caption{Spatial trajectories of an OUP-driven angle without additional torque ($\Omega=0$)
  at constant $n=2$, $D_{\xi}=0.01$, and  $v_0=0.5$  for different correlation times $\tau_c$
  and the time length $t_l=1000$. We see a slight growth in the effective
  displacement for the larger correlation times as also presented in Fig. \ref{tcorrdeff}. }
\label{fortraj2}
\end{figure}
This all is in  perfect accordance to the numerical results, as depicted in Fig.
\ref{tcorrdeff}.  We notice the strong qualitative changes in the
motion of agents which are induced by different $n$.  While for $n=0$
 and $n=1$, the effective displacement decreases with the
correlation time, it even increases with $\tau_c$ for $n=2$. That the
diffusion coefficient increases unbounded with $\propto \sqrt{\tau_c}$
can be seen also in Eq. \eqref{deffoup} where, as mentioned in the last
paragraph above, the integrand increases monotonously with $\Psi_0$, which is
the only parameter dependent on $\tau_c$ for $n=2$.

On the other hand, regarding the limit $\tau_c\rightarrow 0$,
$D_{\text{eff}}$ diverges for $n<2$. Only for $n=2$, the diffusion
coefficient remains finite and converges to the result of Mikhailov
and Meink\"ohn \cite{mikhailov}.
We notice that the effective noise intensity to which our particles
are subjected (i.e., $D_{\xi}/\tau_{c}^n$) increases with $n$ for
correlation times smaller than one, and decreases with $n$ for
correlation times larger than one. For $\tau_c=1$ the diffusion
coefficient of the three cases merge to the same value. In case of
$n=1,2$ for smaller correlation times an effectively larger noise
changes the orientation of the particles repeatedly compared to $n=0$.
That is why the diffusion coefficient is reduced and is largest
for $n=0$. On the other hand, the effective noise intensity acts in the
opposite way for $\tau_c>1$ leading to a reduced  diffusion in case of
$n=0$.

Figure \ref{traj} depicts for $n=1$ different trajectories for a vanishing torque
and $n=1$ within the $\tau_c$ region where $D_{\text{eff}}$ changes
only marginally (see  Fig. \ref{tcorrdeff}). Thus, we recognize
qualitative changes of the motion, which are just induced by the
growth in correlation time, while the variance of the OUP  [Eq.
(\ref{gyrotcorr})] stays constant. One sees that the structure of the
trajectory changes from an uncorrelated random sequence to a waltzing
like motion where longer straight pieces of the trajectory are
interrupted by left- and right-turning pirouettes. We can interpret this
behavior as an increment of the trajectories' persistence as $\tau_{c}$ increases.

Figure \ref{fortraj2} depicts trajectories for $n=2$, while $\Omega=0$
still holds.  Conforming to the behavior shown in Fig.
\ref{tcorrdeff}, the displacement slightly increases for the larger
correlation time, due to the decreasing variance [see Eq.
(\ref{gyrotcorr})], which induces a straighter motion. Therefore,  the
curves also appear smoother for increasing $\tau_c$.

\subsection{Phenomena of applying  finite torque: $\Omega \ne 0$}
\begin{figure}[t]
\begin{center}
\includegraphics[width=0.95\linewidth]{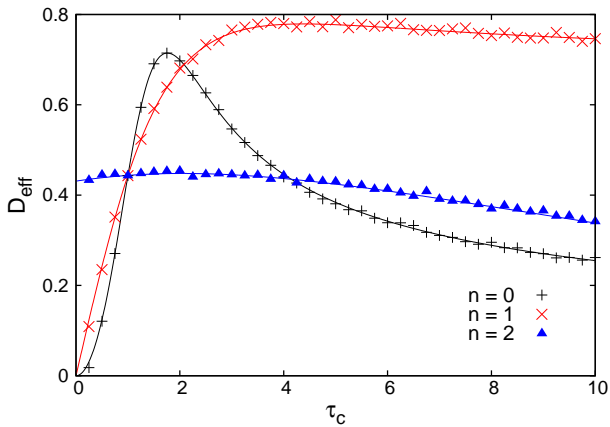}
\end{center}
\vspace{-10pt}
\caption{(Color online) Effective diffusion coefficient $D_{\text{eff}}$ versus
  correlation time $\tau_c$ for several $n$-values, a constant
  torque $\Omega=0.1$, the velocity $v_0=0.5$ and $D_{\xi}=0.01$. Simulations (points)
  agree well with the theory (lines).}
\label{fordeff}
\end{figure}

We next turn to the effects arising from the application of  a constant torque. The
effective diffusion coefficients are displayed in Fig.  \ref{fordeff}
for different $n$-values, together with simulation results as a
function of increasing color $\tau_c$.  We observe that the time correlation induces upon
varying $\tau_c$ a maximal mean square displacement at a finite value
of the correlation time for all three cases. A similar graph could be
also displayed for the dependence on the noise intensity $D_\xi$ (not shown, see also
below).

\begin{figure}[t]
\begin{center}
\includegraphics[width=0.49\linewidth]{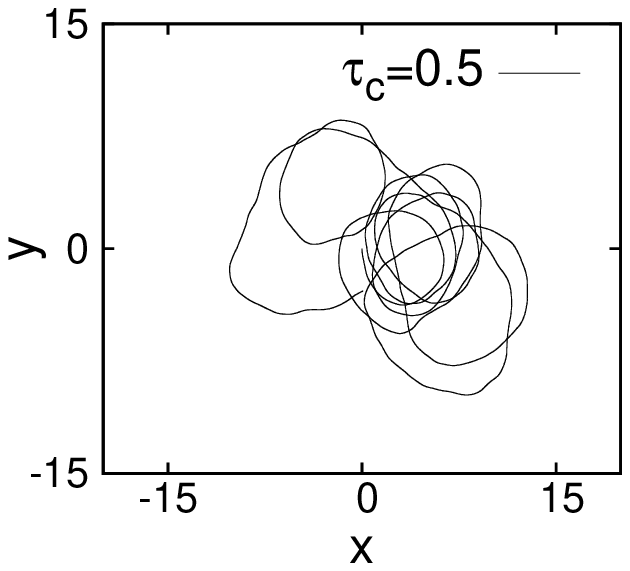}
\includegraphics[width=0.49\linewidth]{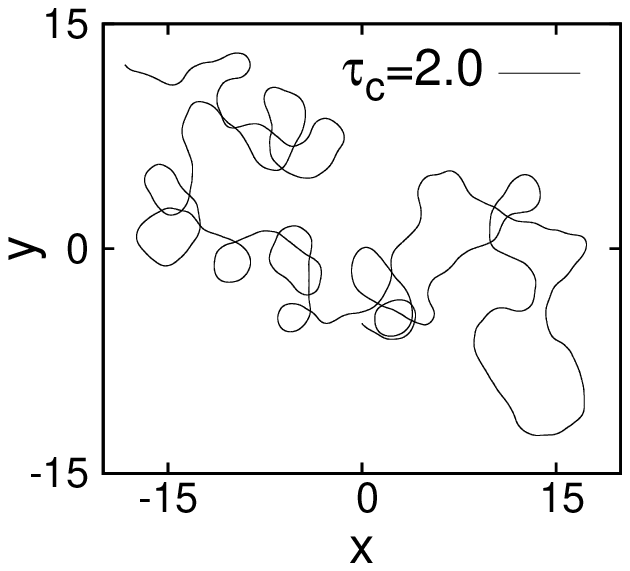}
\includegraphics[width=0.49\linewidth]{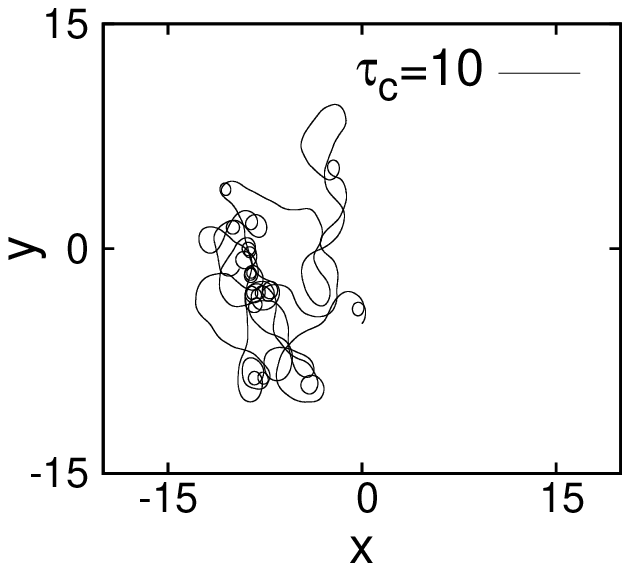}
\end{center}
\vspace{-10pt}
\caption{Spatial trajectories of an OUP driven angle with an
  additional angular force $\Omega=0.1$, $D_{\xi}=0.01$, $v_0=0.5$ and
  with $n=0$, for different correlation times $\tau_c$ and the time
  length $t_l=500$; $n=0$ is used in order to obtain the largest
  $\tau_c$ dependence of the displayed graphs (cf.
  Fig. \ref{fordeff}).}
\label{fortraj}
\end{figure}

In case that $n=0$, where the integrated noise intensity in the OUP dynamics becomes
independent of  the color $\tau_c$, the maximum is most pronounced.
Corresponding trajectories with constant torque are shown in Fig.
\ref{fortraj}. The curvature of these trajectories reflects the
combined influence of the constant torque and of the temporal
correlated random torque forces inherent in $\theta(t)$.

For small color $\tau_c$ the motion is dominantly circular with small random
perturbations. This can be understood in view of Eq.  (\ref{gyrosys}),
since small correlation times imply strong relaxation so that random changes become essentially white noise with
\begin{align}
  \label{white_limit}
  \theta(t)\, \sim \, \sqrt{2D_\xi \tau_c^{2-n}} \xi(t) \;.
\end{align}
Consequently, in the cases that $n=0$ and $n=1$ the white noise sources tend to
vanish. The trajectories follow the motion with the constant
torque $\Omega$ along trajectories with a fixed curvature $\propto
1/\Omega $ and therefore $D_\text{eff}$ decreases.  In contrast to
this behavior, Eq. (\ref{white_limit}) shows that we recover for $n=2$
a white Gaussian angle drive in the limit of small correlation times.

For moderate correlation times a modified curled structure emerges.
For  the stronger correlated systems one observes longer stays at a
certain curvature $\theta$ mixed with longer straightening segments
where the random torque and the constant torque compensate each other.
Thus we retrieve the pirouettes, but now with the preferred
curling orientation of the applied constant torques. Thereby the
straight segments interrupted by the strongly curling pirouettes give
rise to the maximal diffusive behavior and the resonance like
structure shown with Fig. \ref{fordeff}.

Eventually, for very large correlation times $\tau_c$  we find a decrease of the
effective diffusion coefficient, much like  in the situation without
torque [see Fig.  (\ref{tcorrdeff})].  $D_\text{eff}$ converges to zero
for $n=0$ and to a non-vanishing constant for $n=1$.  The case $n=2$
leads [i.e., due to Eq.  (\ref{gyrosys})] to a dominant torque in the
limit of large $\tau_c$, so that the effective diffusion coefficient
decreases toward $0$.

These asymptotic behaviors for small (i.e. $\Psi_0 \rightarrow 0$) and large correlation times
(i.e. $\Psi_0 \rightarrow \infty$)  in Fig.
\ref{fordeff} can be discussed analytically in greater detail:  According to Eqs.
(\ref{eq:deff_as}) and (\ref{eq:deff_as1}) we arrive at the following limiting behaviors:
\begin{align}
  \mathrm{if} \ \ \Psi_0\; \ll & \;1 \quad \rightarrow \quad
  D_{\text{eff}}\;\approx \frac{v_0^4\tau_c^{n-2}}{2 D_\xi}\frac{1}{1+\left(\frac{v_{0}^{4}\Omega}{\tau_{c}^{2-n}D_{\xi}}\right)^2} ,
  \label{deff_omtc1}\\
\mathrm{if} \ \ \Psi_0\; \gg & \;1 \quad \rightarrow \quad
D_{\text{eff}}\;\approx\; \frac{v_0^3}{2}\sqrt{\frac{\tau_c^{n-1}}{D_\xi}}e^{-\frac{v_{0}^{6}\tau_{c}^{n-1}\Omega^{2}}{2D_{\xi}}
} .
\label{deff_omtc2}
\end{align}
Thus, we find  a vanishing effective diffusion
coefficient for correlation times approaching zero if $n \in \{0,1\}$
and one converging to $D_\text{eff}^{\text{white}}$, if $n=2$.  For
large correlation times $\tau_c$, on the other hand, we find with $\Psi \gg 1$   that $D_\text{eff}^{\text{OUP}}$
converges toward zero if $n=2$ or $n=0$.
In distinct contrast, $D_\text{eff}$ approaches a
constant value for $n=1$.

\begin{figure}[tb]
\begin{center}
\includegraphics[width=0.95\linewidth]{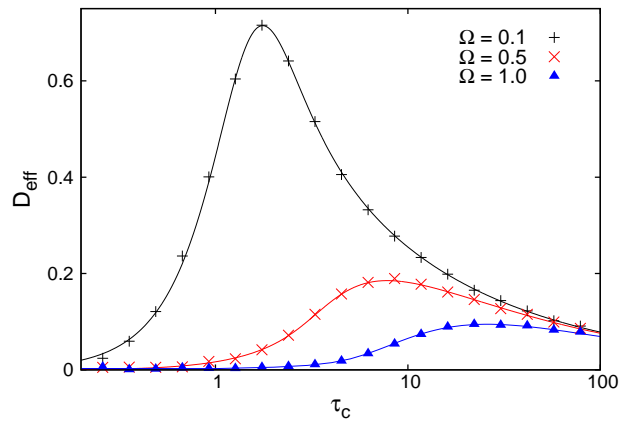}
\end{center}
\vspace{-10pt}
\caption{(Color online) Effective diffusion coefficient $D_{\text{eff}}$ versus
  correlation time $\tau_c$ for different torques $\Omega$
  at $D_\xi=0.01$, $v_0=0.5$ and $n=0$, within simulations
  (points) and in theory [eq. (\ref{msd}); lines].}
\label{phioupfordeff}
\end{figure}

In comparison to the asymptotic behavior without constant torque [cf.
Eqs. (\ref{deff_tc1}) and (\ref{deff_tc2})], we
notice additional multipliers with dependence on $\Omega$ on the
right side of Eqs. (\ref{deff_omtc1}) and (\ref{deff_omtc2}). Both
multipliers map onto the interval $(0,1)$. Thus, the effective
diffusion coefficient always decreases with a constant torque $\Omega
\ne 0$. 
This is depicted in Fig. \ref{phioupfordeff} for $n=0$.
As both limits for large and small correlation times fall to zero, a maximal value for $D_\text{eff}$ occurs,
which not only decreases for growing $\Omega$ but also shifts to larger $\tau_c$ .

\begin{figure}[t]
\begin{center}
\includegraphics[width=0.95\linewidth]{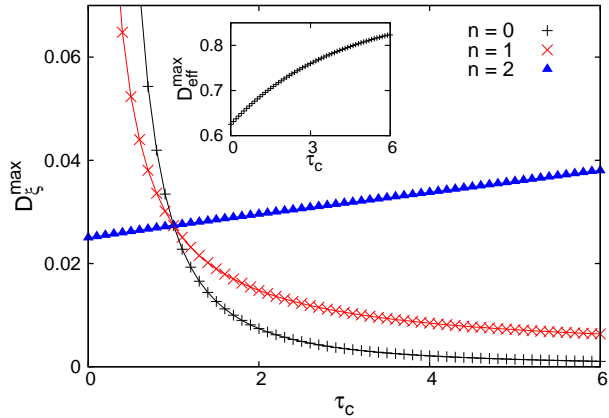}
\end{center}
\vspace{-10pt}
\caption{(Color online) Noise intensity $D_\xi^{\text{max}}$ which maximizes the effective
  diffusion constant for a given correlation time for different $n$ at the velocity
  $v_0=0.5$ and the torque $\Omega=0.1$. The inset shows the maximal $D_\text{eff}$ for these
  parameter pairs, which does not depend on $n$.}
\label{dxi_tauc}
\end{figure}

We now inspect how noise and correlation time influence
each other for different $n$. We have seen
that the effective diffusion constant exhibits a maximum for a finite
non-zero correlation time. Likewise, it has a maximum for a finite
non-zero noise intensity, as seen from the results with white noise (see Fig. \ref{phiwhitefordeff}).  
If we follow the ridge line of the
maxima for fixed $\Omega$ in the $(\tau_{c},D_{\xi})$ parameter space, we obtain
functional dependencies for different $n$-values, as
depicted with Fig.  \ref{dxi_tauc}.  The inset shows the height at the
ridge line (i.e., the value for the maximal effective diffusion
coefficient as function of the correlation time).

For small $\Omega$, we find that the optimal noise intensity
$D_{\xi}^{\text{max}}$, which maximizes the effective diffusion
coefficient at a given correlation time, behaves according to
$D_{\xi}^{\text{max}}\propto \tau_{c}^{n-2}$.  Further analysis shows
that for increased $\Omega$ this relation gradually shifts to
$D_{\xi}^{\text{max}}\propto \tau_{c}^{n-1-\epsilon}$, with a small
finite $\epsilon$. This shift can be justified by the denominator in
Eq.  (\ref{seriesdimless}), in which the second contribution dominates
for larger $\Omega$. Figure \ref{dxi_tauc} depicts this dependence for an
intermediate value of $\Omega$, since
we observe a roughly linear, but still sub-linear, connection for $n=2$,
which converges to the value given for a white angle drive as
$\tau_c\rightarrow 0$. For $n=0$ and $n=1$, we recognize the expected
reciprocal connection.

If we follow the ridge line, at a given correlation time each has the
same height for different $n$. We can understand this by considering
that the parameters $\Psi_0$, $\omega_c$, and $D_c$, which contribute to the
effective diffusion coefficient in Eq. (\ref{deffoup}), are all
containing the same factor $D_{\xi}\tau_c^{-n}$ as sole $n$ dependence 
[cf. Eqs. (\ref{eq:psi0}) and (\ref{dcwc})]. Hence, according
to the discussion above, $D_\text{eff}$ does not depend on $n$ anymore while regarding $D_{\xi}^{\text{max}}$.

\section{Conclusions}

With this work the stochastic dynamics of  active particles with a constant speed
that are  additionally driven by an overall fluctuating torque has been studied.
The cases of correlated and uncorrelated angle dynamics were considered and
analytical results for the mean square displacement  were derived.  We
discussed in this context the dependencies on characteristic
parameters of our system, such as the correlation time $\tau_c$ and the
corresponding noise intensity $D_{\xi}$, to find maxima in the mean square displacement. With respect
to the correlation time and to the noise intensity a maximum of the effective
diffusion coefficient $D_{eff}$ was identified. We also gave a qualitative
explanation of the maximal spreading of the agents by an inspection of
the individual trajectories. The random composites of stretched and
curved parts in the trajectories can give rise to an increased mean
square displacement.

It was also shown that the constant torque decreases the mean square
displacement. Therefore, the persistence of motion and presence of a permanent
torque are two appropriate instruments to optimize the two-dimensional
motion of active agents.

In view of the fact, that the displacement of an agent forms a strong
link between theoretical and experimental studies of active particles,
we hope that further work within this field benefits from our
theoretical reasoning presented here.

Identifying the trajectories of our studied dynamics with the
structure of a macromolecule, spin-off applications in the field of polymer
physics are conceivable. Especially, the white Gaussian angle drive
without a torque has  its polymer analog in form of the famous
worm-like chain model \cite{doi}. Moreover, the similarities between
the paths in Fig. \ref{fortraj} and Fig. \ref{traj} and correlated animal
movements are striking for the case of animals, whose angle
dynamics is determined by the past states.

Finally, a useful extension of our model would be the adaptation to more realistic
biological systems by accounting as well for an additional velocity dynamics $v(t)$.
Simple velocity models, wherein one can decouple
the $v$- and $\phi$-dynamics, are leading thereby directly to the
discussion in \cite{peruani}, exhibiting a multi-crossover structure of the
mean square displacement, which is due to the different time scales within
those systems.

\acknowledgments
\noindent This work has been supported by the VW
Foundation via project I/83903 (L.S.-G.) and, as well,  I/83902 (P.H.).
P.H. also acknowledges the support by the DFG
excellence cluster ``Nanosystems Initiative Munich'' (NIM).

\end{document}